\documentclass[conference, a4paper]{IEEEtran}
\IEEEoverridecommandlockouts

\usepackage{amsmath}
\usepackage{balance}
\usepackage{cite}
\usepackage[T1]{fontenc}
\usepackage{inconsolata}
\usepackage{listings}
\usepackage{url}

\newcommand{\src}[1]{\texttt{\small#1}}
\usepackage[absolute]{textpos}
\usepackage[hidelinks]{hyperref}

\begin{document}

\begin{textblock}{15}(0.5,14.8)
{
\noindent\hrulefill

\noindent\fontsize{8pt}{8pt}\selectfont\copyright\ 2019 IEEE. Personal use of this material is permitted. Permission from IEEE must be obtained for all other uses, in any current or future media, including reprinting/republishing this material for advertising or promotional purposes, creating new collective works, for resale or redistribution to servers or lists, or reuse of any copyrighted component of this work in other works. \hspace{5pt} This is the accepted version of: M. Sul\'ir, J. Porub\"an. Designing voice-controllable APIs. 2019 IEEE 15th International Scientific Conference on Informatics, IEEE, 2019, pp. 393--398. \url{http://doi.org/10.1109/Informatics47936.2019.9119302}

}
\end{textblock}

\title{Designing Voice-Controllable APIs}

\author{\IEEEauthorblockN{Mat\'u\v{s} Sul\'ir, Jaroslav Porub\"an}
\IEEEauthorblockA{\textit{Department of Computers and Informatics} \\
\textit{Faculty of Electrical Engineering and Informatics} \\
\textit{Technical University of Ko\v{s}ice}\\
Ko\v{s}ice, Slovakia \\
\{matus.sulir, jaroslav.poruban\}@tuke.sk}
}

\maketitle

\begin{abstract}
The main purpose of a voice command system is to process a sentence in natural language and perform the corresponding action. Although there exist many approaches to map sentences to API (application programming interface) calls, this mapping is usually performed after the API is already implemented, possibly by other programmers. In this paper, we describe how the API developer can use patterns to map sentences to API calls by utilizing the similarities between names and types in the sentences and the API. In the cases when the mapping is not straightforward, we suggest the usage of suitable annotations (attribute-oriented programming).
\end{abstract}


\section{Introduction}

Voice control is slowly becoming a mainstream form of human-computer interaction. For this reason, developers often integrate voice control into the applications being developed. In this article, we will focus on simple, one-sentence commands after which we expect the computer to perform the desired action. Such a process consists of two main parts \cite{Rogowski12industrially}: the translation of the voice input to a natural language sentence and the selection of an appropriate action according to the meaning of the sentence. In this paper, we suppose the voice recognition is implemented by a third-party library or service. Therefore, we will focus on the second part: the mapping of a sentence to the desired action.

Since we are interested in the implementation of the voice control from the developer's view, by an action we mean a call to an appropriate method in an API (application programming interface) with correct parameters. There already exist many approaches to map a natural language sentence to API calls. Some of them utilize probabilistic grammars and heuristics \cite{Gvero15synthesizing}, others perceive phrases only as lists of keywords \cite{Little06translating}. An increasingly popular approach is the application of machine learning to translate natural language sentences to API snippets \cite{Nguyen16t2api}.

The common sign of these approaches is that they perceive voice control only as an afterthought -- when the API is already implemented, probably by other developers. In contrast to them, we will show how the API can be tailored to be voice-controllable, predominantly by exploiting the similarity of names and data types used in the sentences and the API. For example, a sentence ``open the door for 5 seconds'' could be interpreted as a call to \src{door.open(5)} during the execution of a program, provided there is a method \src{void open(int seconds)} in the class \src{Door}. When such similarities are not observed, we suggest utilizing annotations (attribute-oriented programming) to specify the necessary details of the sentence-to-API mapping.

Throughout the paper, we also discuss how such APIs could be analyzed by a voice command recognition framework, which would then execute an appropriate action given a natural language sentence. A prototype implementation of such a framework, supporting many of the described features, is available online\footnote{\url{https://github.com/sulir/voice-control-demo}}.

\section{Mapping Patterns}

In this section, we will gradually describe individual patterns of mapping between a natural language sentence and an API call. Our examples will be related mainly to the hardware device control domain, but it is not restricted to it. Although we will use Java, the approach is generally applicable to any object-oriented programming language with sufficient metaprogramming features, static typing is also helpful for some patterns.

\subsection{Verbatim Name Mapping}

Suppose we would like to turn on light in a room after saying ``turn on light''. The most primitive way to map this voice command to a method in the API would be to name the method \src{turnOnLight} and annotate it with a marker annotation \src{VoiceControllable}, so our framework could recognize it as a voice-controllable function:
\begin{lstlisting}
@VoiceControllable
public static void turnOnLight() {
    // implementation turning on the light...
}
\end{lstlisting}
The name of the method would be split into individual words and transformed to lowercase. Using exact string matching, the input ``turn on light'' could be then mapped to this method, which would be executed without any parameters.

Of course, having one class with many static methods is a bad programming practice. Therefore, suppose we have separate classes to control lights, monitors, speakers, and other devices. For simplicity, let us say these classes are named \src{LightService}, \src{MonitorService}, etc. In modern application frameworks, service constructors are usually not called manually -- the framework initializes them using techniques such as dependency injection. Thus we will assume we have an instance of every necessary class readily available and we will focus on calling methods of this object. We can transform the previous example to this code:
\begin{lstlisting}
@VoiceControllable
public class LightService {
    public void turnOn() {
        // implementation turning on the light...
    }
}
\end{lstlisting}
The annotation over a class means all its methods are now ``voice-controllable''. Our framework searches for all such classes and their methods, building potential command-to-method mappings. Since \src{Service} is an implementation-oriented name irrelevant to the problem domain \cite{Kollar12abstraction}, our framework strips it from the class name when scanning it. The rest of the class name is appended to the method names. Therefore, the \src{turnOn} method of class \src{LightService} will be called when the command ``turn on light'' is issued.

\subsection{Minor Variations}

Since natural language is not restricted to exact phrases, we must count with some variations in the commands. First, we should stem all words in both the command and the class/method names using an appropriate stemmer. This way, we allow for variation in individual words since only their root forms are compared. For example, the command ``turn on lights'' can now execute the aforementioned method, even if ``Light'' is singular in the class name.

Next, we should allow permutations of the words. For example, ``turn on lights'' can be expressed also as ``turn lights on''. In general, we need to find a compromise between allowing all permutations (which allows diverse sentences at the price of higher ambiguity) and supporting only certain kinds of permutations.

We can also allow extraneous words to be present in the command. Typical examples include stop words (``the'' in ``turn on the light'') or courtesy phrases (``please turn on the light''). An extreme example is the sentence ``Please be so kind and turn on the light, dear.'' As long as the extraneous words are not contained in the names in our API, we can safely remove them from the sentence without any ambiguity. To enable more precise decisions, we can annotate the method with words we expect to encounter in the command. For example, we would like to execute the method \src{turnOff} by the command ``turn off all lights'':
\begin{lstlisting}
@ExtraWord("all")
public void turnOff() { ... }
\end{lstlisting}

A similar case is the non-presence of certain words from API names in the natural language command. For instance, the command ``light off'' does not contain the word ``turn'' from the method \src{LightService::turnOff}. To help the mapping system to make a correct decision, we can annotate the method with \src{@OptionalWord("turn")}.

By allowing all of the mentioned variations at once, we can achieve a system which is highly flexible and accepts a wide range of different sentences -- at the price of potential ambiguity. Essentially, the framework should find a method in the API which has the highest number of words common with the input. When there are ties or the number of common words is too low, we can start taking the \src{ExtraWord} and \src{OptionalWord} annotations into account. If the situation is still indecisive, the voice assistant can simply list the most probable options and ask the person to select one of them.

\subsection{Parameters}

Now we will take into account methods with parameters. The simplest case is a method with one numeric parameter:
\begin{lstlisting}
public void turnOn(int number) { ... }
\end{lstlisting}
Typical commands to execute this method are: ``turn on light number 1'' and ``turn on light 1''. In the first sentence, the argument value follows the parameter name ``number'', which means the argument can be matched by its name. In the second case, the parameter is matched only by its type -- the input sentence contains a number and the API method has a numerical parameter.

Mapping enum-typed parameters is straightforward too. Suppose \src{Position} is defined as an enumeration:
\begin{lstlisting}
public enum Position {
    LEFT, MIDDLE, RIGHT
}
\end{lstlisting}
We also have a method with a \src{Position} parameter:
\begin{lstlisting}
public void turnOn(Position position) { ... }
\end{lstlisting}
Then this method can be called by the command ``turn on the left light''. On the other hand, mapping of string parameters is more problematic:
\begin{lstlisting}
public void turnOn(String name) { ... }
\end{lstlisting}
If the user says ``turn on the light named `front'{}'', the system can use the matching based on parameter names. However, mapping the sentence ``turn on the front light'' to this method is questionable in general, since we do not have any way to enumerate all possible valid values of a \src{String} parameter. A potentially dangerous way to determine if this method matches the command is trying to execute it and observing if it does not throw an exception. There exist better ideas, though. If the list of valid values is fixed at a compile time, it can be supplied directly:
\begin{lstlisting}
public void turnOn(
    @Values({"front", "back"}) String name
) { ... }
\end{lstlisting}
In such cases, however, it is much better to use enumerations instead. Most often, the set of valid values can be enumerated only at runtime. Then we can create a class with a method returning these values:
\begin{lstlisting}
public class LightNames implements ValueSet<String> {
    public Set<String> getValues() {
        return configuration.getValidLightNames(...);
    }
}
\end{lstlisting}
Subsequently, we connect this list with a parameter using an annotation:
\begin{lstlisting}
public void turnOn(
    @ValidValues(LightNames.class) String name
) { ... }
\end{lstlisting}

Now we will consider parameters of any class in general, e.g., a parameter of type \src{Color}:
\begin{lstlisting}
public void setColor(Color color) { ... }
\end{lstlisting}
In the command ``set light color to green'', we do not know how to map ``green'' to \src{Color(0, 255, 0)}. It is possible that \src{Color} has a constructor accepting a string, in which case we can utilize it. However, this still does not solve the problem of the enumeration of all valid values described in the previous example. Therefore, we suggest using a string-to-object mapping via annotations. Each parameter of a custom class should be annotated with a mapper:
\begin{lstlisting}
public void setColor(
    @StringMapping(ColorMapper.class) Color color
) { ... }
\end{lstlisting}
Then the \src{ColorMapper} class can look like this:
\begin{lstlisting}
public class ColorMapper
implements StringMapper<Color> {
    public Map<String, Color> getMap() {
        return Map.of(
            "red", new Color(255, 255, 0),
            "green", new Color(0, 255, 0),
            ...
        );
    }
}
\end{lstlisting}
This way, we can both enumerate all possible values of a \src{Color}-typed parameter and map a string such as ``green'' to a \src{Color} object. The approach is not limited to a constant associative array of string--color pairs: the mapping could be also dynamically generated by arbitrarily complicated code. Note that since the mapping is specified in a separate class, we did not modify the original \src{Color} class in any way -- it can be defined in a third-party library without problems.

Next, we consider a method with multiple parameters:
\begin{lstlisting}
public void setColor(int number, Color color) { ... }
\end{lstlisting}
As long as the parameters are of distinct types and the sets of valid values of all types are disjunctive, we can map various commands to a method execution without too much ambiguity: e.g., ``set light 3 to blue'', ``I would like yellow color for light 4''. However, methods with multiple parameters of same or similar type are more problematic:
\begin{lstlisting}
public void setBrightness(int light, double brightness)
{ ... }
\end{lstlisting}
Here, ``set light 1 to brightness 50'' can be mapped to a correct execution using the parameter names. However, the sentence ``set brightness of light number 2 to 30'' could be probably mapped only by the position of arguments, which would not work correctly if the parameters were switched in the API.

Note that parameters have an important role in the process of the API method selection. Since it is not valid to call a method without all mandatory (non-null) arguments filled, they must be all present in the input sentence. For example, if the command does not contain any number, we can assume it does not match any method with a numeric parameter.

\subsection{Collections}

Suppose we want to dim multiple lights at once. A collection, such as a set, is an ideal candidate for this:
\begin{lstlisting}
public void dim(Set<Integer> lights) { ... }
\end{lstlisting}
The content of the parameter \src{lights} can be expressed by enumerating all values or by specifying a range: ``dim lights 1, 7 and 9'' or ``dim lights 6 to 10''. In case the expected range can be very large, we recommend using a lazy collection denoted by an interface, such as \src{Iterable<Integer>}.

Similarly to numeric collections, we accept collections of enumerations or other classes:
\begin{lstlisting}
public void dim(Set<Position> lights) { ... }
\end{lstlisting}
Naturally, ranges are not supported in such cases. We can still fill this set with values by naming them all, though: ``dim the left and middle light''. Alternatively, commands such as ``dim all lights'' should work if there is a way to enumerate all valid values of the parameter programmatically.

\subsection{Synonyms}

Many words have synonyms which can a user utter instead of the words used in the API names. A natural way to cope with such situations is the usage of thesauri and lexical databases such as WordNet \cite{Miller95wordnet}. Any word in the API can be then replaced by its synonym to perform a successful match with the natural language command. However, consider the following example of a method in class \src{ScreenService} (controlling a multi-monitor ambient user interface \cite{Galko16tools}), where \src{State} is an enumeration with values \src{ON} and \src{OFF}:
\begin{lstlisting}
public void set(int screen, State state) { ... }
\end{lstlisting}
We would certainly like to say not only ``set screen 1 to `on'{}'' but also ``turn screen 1 on''. However, ``turn'' is not a synonym of ``set'' in general -- only in this specific case. Therefore, we devised a way to specify method-local synonyms via annotations:
\begin{lstlisting}
@Synonym(of = "set", is = "turn")
public void set(int screen, State state) { ... }
\end{lstlisting}
This synonym can be applied only during the matching of this particular method. Analogously, we support package-local, class-local and parameter-local synonyms using annotations over packages (in a special file \src{package-info.java}), classes, and parameters, respectively. It is also allowed to specify multiple synonyms per element, e.g.:
\begin{lstlisting}
@Synonym(of = "screen", is = {"display", "monitor"})
@Synonym(of = "turn", is = "switch")
public class Screen { ... }
\end{lstlisting}

\subsection{Fallback}
\label{s:fallback}

If we encounter difficulties during the application of the aforementioned mapping patterns, we can always specify the voice commands manually. For example, suppose we have a class \src{SpeechService} with a method \src{pronounce(String sentence)} which says out loud the given sentence. Since the sentence can be completely arbitrary, matching will likely fail. Therefore, we specify the voice command as a regular expression:
\begin{lstlisting}
@VoiceCommand("say (.*)")
public void pronounce(String sentence) { ... }
\end{lstlisting}
The content of the group in the parentheses will be supplied as a parameter value. Because a manually specified command has a high priority, we can successfully match commands such as ``say `I like turning off the screens'{}'' even when they contain words present in the API.

\section{Mapping Process}

In our prototype implementation\footnote{\url{https://github.com/sulir/voice-control-demo}}, we decided to use a simple sentence recognition algorithm designed to allow for relatively large deviations of the input sentences from the prescribed forms. Now we will briefly describe it.

First, the input sentence is matched against all fallback regular expressions (section~\ref{s:fallback}). If a match is found, the process is stopped and the annotated method is executed.

Next, for each voice-controllable method, we try to type-match all its parameters with the sentence. For example, if a method has a numeric parameter and a \src{Color} enumeration parameter, the sentence is searched for a numeral and a word denoting a color. The result is a list of potentially matching methods, along with word-to-parameter mappings.

For each method in this list, a score of similarity with the input sentence is calculated: Let $W_M$ be the set of words contained in the class and method name. Let $W_S$ be the set of words contained in the input sentence, excluding the parameter values matched in the previous steps. The score is computed as the Jaccard index \cite{Jaccard12distribution} of these two sets:
\[
\frac{|W_M \cap W_S|}{|W_M \cup W_S|}
\]
If the class or method is annotated by synonyms, multiple variants of the set $W_M$ are constructed, with the words in the identifiers gradually replaced by their corresponding synonyms. The resulting score is a maximum of the scores computed for individual variants.

Finally, the method with the highest score is executed. If multiple methods have the same score or if no method receives a score higher than a certain threshold (e.g., 0.2), the user should be asked to reformulate the command.

It is important to note that the described algorithm is not the only one possible. The ideas described in the previous section are compatible, for instance, with strict regex-based solutions based on the precise matching of sentences with regular expressions generated from the identifier names and annotations. Another possibility is a grammar-based approach, which would distinguish parts of sentences: Some of them could be derived automatically if possible, the rest will be determined by manually written annotations.

\section{Related Work}

Many approaches mapping sentences to API calls are grouped under the umbrella term of program synthesis, particularly program synthesis using natural language input \cite{Gulwani10dimensions}. Desai et al.\cite{Desai16program} synthesize source code written in various domain-specific languages, thanks to a dataset of sentence--code pairs used for training. Gvero and Kuncak \cite{Gvero15synthesizing} synthesize Java expressions using probabilistic grammars and heuristics. In T2API, Nguyen et al. \cite{Nguyen16t2api} perceive program synthesis as a statistical translation process from natural language to a programming language. Little and Miller \cite{Little07keyword} generate Java code from a set of brief keywords. Some synthesis approaches are focused on particular domains or technologies: e.g., SQL query generation \cite{Yaghmazadeh17sqlizer}, smartphone automation script synthesis \cite{Le13smartsynth}, bot API invocations \cite{Zamanirad17programming}. All of the mentioned approaches perceive APIs as black boxes, which are already designed. In contrast to them, our idea is to engage the API designers in the process of natural language command specification.

Landh\"{a}u{\ss}er et al. \cite{Landhausser17nlci} designed NLCI (natural language command interpreter), which has a goal similar to ours -- to perform an action in the API, given a natural language sentence as an input. In contrast to us, they first transform the API into an ontology by analyzing the relationships between elements in the code and combining it with a general-purpose ontology. Furthermore, they do not support simple mapping customization via annotations.

The command execution approach by Little and Miller \cite{Little06translating} is based on the similarity of names used in the sentence and the API. They also perceive sentences as lists of keywords, allowing for variations such as extraneous words. However, they do not allow any customization using annotations, since they consider APIs to be developed by a third party and thus not modifiable.

Naturalistic programming \cite{Pulido-Prieto17survey} is a paradigm aiming to make the source code look more like natural language. For example, Kn\"oll et al. \cite{Knoll11naturalistic} discuss naturalistic types which include the mapping of natural language quantities such as ``nearly all'' to exact numeric intervals. Compared to them, we aim to integrate voice control with existing, traditional programming languages instead of designing new ones.

There exist guidelines on how to design APIs in general \cite{Bloch06how} and an overview of design decisions to be made when creating an API \cite{Stylos07mapping}. Nevertheless, none of these works take voice-controllability into account.

Hirzel et al. \cite{Hirzel17i} describe an idea of grammars for dialog systems, including virtual voice assistants. However, they do not try to solve the problem of the mapping of sentences to API calls.

Commercial system APIs, such as Google Voice Actions for Android\footnote{https://developers.google.com/voice-actions/custom-actions} or SiriKit \footnote{https://developer.apple.com/documentation/sirikit} are often limited to certain domains and action types. Furthermore, they require some effort to integrate, such as the creation of configuration files or implementation of non-trivial interfaces.

YAJCo \cite{Poruban09annotation} is a parser generator utilizing the similarity between the relations of program elements in Java source files and production rules of computer language grammars. The core ideas behind this article stemmed from YAJCo, however, this time they are applied to natural languages.

\section{Conclusion and Future Work}

In this paper, we described patterns of mapping between a natural language command and an API method call. These patterns are based on:
\begin{itemize}
\item class, method, and parameter names,
\item parameter types and positions.
\end{itemize}
The mapping can be enabled simply by placing the annotation \src{@VoiceControllable} over a class or a method. When these natural mapping patterns are not sufficient, the programmer can adjust the mapping by using annotations, such as:
\begin{itemize}
\item \src{@StringMapping(Mapper.class)} to specify the string-to-object mapping for a parameter,
\item \src{@ValidValues(ValueSet.class)} to enumerate possible values of a parameter at runtime,
\item \src{@Synonym(of="word1", is="word2")} over packages, classes, methods, and parameters to specify local synonyms,
\item \src{@VoiceCommand("regex (param)")} to specify an exact regular expression whose match will execute the given method.
\end{itemize}

There are many limitations of the described work. First of all, we did not yet perform full validation of our ideas. We should create a golden standard of sentence-to-API mappings (or utilize an existing one). Then we need to validate our approach by measuring the accuracy of the algorithms based on our ideas when compared to the golden standard. We hypothesize our simple approach based on word similarities would work well for small or medium-sized APIs, but it could be problematic for larger ones.

Next, we described only a small portion of all useful patterns. In the future, we could devise more elaborate ways to express numerical ranges, binary operators, various collection types, exceptions, etc. String-to-object mappers could be improved too. In addition to the manual definition of synonyms, domain ontologies could be used too.

The examples mentioned in this article are very simple -- in order to show the point of our approach. We should inspect larger APIs from multiple domains and assess the applicability of our patterns to them.

Finally, in this article, we were interested only in simple, one-sentence inputs being mapped to single API calls. The approach could be extended to support nested API calls or more complex natural language dialogs in the future.

\section*{Acknowledgment}

This work was supported by Project VEGA No. 1/0762/19 Interactive pattern-driven language development. This work was also supported by FEI TUKE Grant no. FEI-2018-57 ``Representation of object states in a program facilitating its comprehension''.

\balance
\bibliography{informatics}

\end{document}